# The Detection of Human Spreadsheet Errors by Humans versus Inspection (Auditing) Software

Salvatore Aurigemma
University of Hawaii
SA8@Hawaii.edu

Raymond Panko
University of Hawaii
Panko@Hawaii.edu, Ray@Panko.com

## ABSTRACT

Previous spreadsheet inspection experiments have had human subjects look for seeded errors in spreadsheets. In this study, subjects attempted to find errors in human-developed spreadsheets to avoid the potential artifacts created by error seeding. Human subject success rates were compared to the successful rates for error-flagging by spreadsheet static analysis tools (SSATs) applied to the same spreadsheets. The human error detection results were comparable to those of studies using error seeding. However, Excel Error Check and Spreadsheet Professional were almost useless for correctly flagging natural (human) errors in this study.

## 1 INTRODUCTION

While people are about 95% to 98% accurate when they make spreadsheet cells entries, they are only about 50% to 80% successful when they attempt to detect if there is an error in a cell [Panko, 2010b]. While improving accuracy during development is important, improving yield during error detection is the sine qua non for improving spreadsheet accuracy.

> *While improving accuracy during development is important, improving error detection is the sine qua non for improving spreadsheet accuracy.*

### 1.1 Error Detection Methodologies for spreadsheets

There are several ways to detect errors in spreadsheets, including testing, inspection, and static analysis tools. We do not include "auditing." The purpose of an audit *is not* to find all errors; rather, its job is to sample cases systematically in order for the auditor to develop an opinion of how well a process is controlled. In contrast, the specific purpose of spreadsheet inspection *is* to attempt to find all errors in a spreadsheet. Most spreadsheet "auditing" studies have really been spreadsheet inspection studies. Nor will use the term "debugging." Inspection is designed to detect errors. Debugging, by definition is finding the root cause of an error once an error is detected.

In *testing*, the tester studies the specifications, defines test cases (sets of inputs), enters them into the spreadsheet, and compares the actual spreadsheet calculations to an "oracle" that specifies what the results *should* be. Unfortunately, the only way to develop an oracle for the results of a complex spreadsheet may be to build another spreadsheet because the calculations in a spreadsheet often are too complex for the manual computation. In addition, spreadsheets often have a large number of inputs, so there is a combinatorial explosion in the definition of possible equivalence classes for test cases. Also, while programming languages have good facilities for testing modules in a program, which reduces the combinatorial explosion in test cases, spreadsheets do not have these facilities [Ettma, Janssen, and de Swart, 2001]. With better testing tools, spreadsheet testing may be effective. Without those tools, testing is very difficult and is rarely done at a professional level.

Another way to detect errors in spreadsheets is to *inspect* spreadsheets cell by cell. Code inspection has long been used in software verification and validation. Nearly all code inspection methodologies follow a method created by Fagan [1976]. Although this method was developed primarily by trial and error, it makes strong sense in terms of general human error research. It recognizes that software engineers are only about 40% to 60% successful, on average, in detecting errors [Panko, 2006]. So it requires team inspection. In addition, it limits the number of lines of code that can be inspected in a session to about 100 statements because research has shown that longer inspections produce lower detection rates [Panko, 2006]. For the same reason, preparation and meeting time are each limited to about two hours. Finally, Fagan inspection requires the inspection team members to be extremely familiar with the code and its goals before they inspect it for errors. A round of code inspection done under these conditions usually finds 60% to 80% of all errors [Panko, 2010a]. Consequently, companies do multiple rounds of inspections at various levels of integration.

While spreadsheet testing runs into quite a few problems, spreadsheet inspection following something like the Fagan [1976] methodology appears to be feasible. In the laboratory, individual subjects with only moderate Excel skills and limited inspection training consistently find 50% to 80% of all seeded errors in experimental spreadsheets [Panko, 2010b]. One study [Panko, 1999] showed that three-person team inspection worked about as well as expected compared to individual inspection. Formal studies have not been done on the appropriate number of formulas per inspection or on other details of Fagan inspection or similar methodologies that have been developed over time in programming.

In this paper, we will equate the term "inspection" with inspection that has most of the features of Fagan inspection.

## 1.2 The Cost of Extensive Testing and Inspection

Both testing and inspection are expensive in software development. Software companies typically spend 30% to 40% of their development effort doing testing and inspection. Consequently, the final fault rate (errors in programs are called faults) is about 0.1% to 0.2% of all lines of code at shipping time [Panko, 2010a]—instead of the original 2% to 5% at the unit level [Panko, 2010a]. In contrast, while spreadsheet experiments indicate that the error rate after development is also about 2% to 5%, final error rates in inspections of operational spreadsheets are almost as high [Panko, 2010b]. This indicates that error reduction is not taking place after development, or at least not enough to be effective. This is not surprising because surveys have shown that extensive spreadsheet testing is rare, although cursory spreadsheet testing is common [Caulkins, 2007; Panko, 2010b].

The close parallel between spreadsheet and software error rates during development and inspection should not be surprising. The two are of comparable complexity, and human error research has shown that cognitive processes of comparable complexity have similar error rates [Panko, 2010a].

In effect, spreadsheet testing and inspection are about where software testing and inspection were in the 1960s. At that time, many programmers and software vendors believed that spending 30% to 40% of all resources on testing was not justified. Hard experience showed that it was. Programming errors often create highly visible results such as crashes. In contrast, spreadsheet errors usually only give incorrect numbers—which may not be recognized by users as incorrect. This phenomenon of silent errors has prevented the kind of feedback that we have long had in programming.

Commercial spreadsheet validation and verification companies have found over time that extensive testing and inspection really is necessary. The author's experience with doing commercial inspections of spreadsheets is that the process requires an average of about 50 hours per person (with wide variation) and a team of three to five. Croll [2003] described how commercial inspection is typically done in the City of London. He said that the time taken in a study ranges from 25 hours to several hundred hours. In addition, Ettma, Janssen, and de Swart [2001] said that audits at PricewaterhouseCoopers in Amsterdam typically took 25 to 75 hours for spreadsheets of moderate size.

To give another example, the U.S. Army Corps of Engineers (USACE) does many large environmental projects. Planning analysis is done with custom software or a spreadsheet program. After having legal problems when adversaries inspected their models and found errors, USACE required new models to be inspected by a multidisciplinary team consisting of both domain knowledge experts and tool (software or spreadsheet) professionals. This process requires a few hundred hours.

### 1.3 Spreadsheet Static Analysis Tools (SSATs)

One possibility for spreadsheet error detection is to use *spreadsheet static analysis tools (SSATs)*, such as Spreadsheet Professional, SpACE, and the Operis Toolkit. These programs have a number of functions. One is to look through the spreadsheet and *tag* (mark) cells that appear to be errors or are highly risky. Humans then look at the cells and decide if they actually contain errors. Incidentally, we call these tools instead of programs because while some are indeed programs, others are built into spreadsheet programs themselves. Consequently we used the more general term *tool*.

Another common SSAT function is drawing a map (graphical layout) of the spreadsheet's structure. This can give the inspector a better understanding of the structure of the spreadsheet. In addition, inconsistencies in the visual structure may indicate errors.

The tagging function in spreadsheet static analysis tools is similar to the functions of static analysis programs in software development. Software testing programs are run against software modules in order to flag potential errors or areas of high risks.

In software development, software testing programs are widely used. In the United States, software static analysis vendor Coverity survey found that 88% of all U.S. software vendors were using static analysis programs in software development. However, static analysis programs are not are not replacements for full software testing and inspection. They normally are used only at the unit testing level. Even then, they are only used for prescreening because they can only catch certain types of errors.

An analogy is the use of spell checking in word processing. Spell checkers are very effective at finding typographic errors that result in a non-word text string (e.g.: *then* to *theb*). However, if the typographical error creates a text string that is a word in the dictionary (e.g.: *then* to *them*), the spell checker is worthless. Grammar checking is even more problematic because it generates so many false positives and often gives bad grammatical advice. A study found that students who used grammar checkers produced less grammatical papers than students who did not use grammar checkers [Galletta, et al., 2005]. This happened because students with grammar checkers often took poor advice from the tool. No teacher would ever accept a poorly written paper on the grounds that the student ran a spell checker and grammar checker against it and then did not follow this with real proofreading. No software company would accept a unit that was only tested by a static analysis program.

A number of spreadsheet developers and analysts have argued that extensive spreadsheet testing comparable to good practice software testing is too expensive. They argue this despite the fact that spreadsheet and software error commission and detection rates are very similar. They give no justification other than using terms like "impractical." Similar arguments were made in the 1950s and 1960s, in the early days of software development. They are no longer made in software development.

Although spreadsheet inspection programs might be very useful for error prescreening, some analysts have proposed the use of SSAPs *instead of* doing full inspection or testing [e.g., Nixon and O'Hara, 2001]. This is attractive from a cost point of view; but are SSAPs effective enough to replace inspection? The possibility of replacing testing and inspection with static inspection programs needs to be studied to see if it is safe and effective.

## 2 STUDIES OF SPREADSHEET STATIC ANALYSIS TOOL SAFETY AND EFFECTIVENESS

### 2.1 The Nixon and O'Hara Study

Two fairly large studies have examined the effectiveness of static inspection programs. The first was Nixon and O'Hara [2001].They used a version of an actual corporate spreadsheet. They then seeded the spreadsheet with 17 errors. Nixon used Excel's auditing functions to inspect every cell. He also applied several SSAPs to the spreadsheet—Spreadsheet Detective, Excel Auditor, Operis Analysis Kit, and Spreadsheet Auditing for Customs and Excise (SpACE). Although Excel Auditor was built for cell-by cell error inspection, only its automated tools were used. This was done because Nixon and O'Hara [2001] argued that these tools are best used as ways to reduce cost by replacing cell-by-cell inspection.

Nixon, who seeded the spreadsheet with errors, ran four of the tools against the spreadsheets. (SpACE was used by an experienced SpACE user, the redoubtable Ray Butler). For each seeded error and tool, Nixon rated whether the spreadsheet passed, almost passed (was not entirely convincing that there was an error), nearly failed, or failed. The success metric was the percentage of errors on which the software passed or nearly passed. Operis Analysis Kit scored 65%, while the hobbled Excel Auditor scored 27%. Space achieved 84%, and Spreadsheet Detective achieved 80%.

Overall, the Nixon and O'Hara [2001] study found that some spreadsheet static analysis tools appeared to be about as good as individual human inspectors [e.g, Bishop and McDaid, 2008;Galletta, et al., 2003 and 1996-1997, Howe and Simkin, 2006, McKeever and McDaid, 2009; and Panko, 1999]. This comparability suggests that SSAPs might indeed be able to replace human inspection. Of course, SSAPs might be detecting different errors than developers, so both unaided human inspection and SSAP-aided inspection would be best. Also, while SSAPs detect errors more rapidly than full human inspection, they did not catch all errors, and this would again argue for using both approaches. However, the combination of speed and high detection rate would argue for these tools to be very widely used.

## 2.2 The Problems of Error Seeding

However, it is possible that the high detection error rates were due to the type of errors used to seed the spreadsheet. Error seeding has been used extensively in software engineering [Offutt & Hayes, 1996; Meed & Siu, 1989; Knight and Ammann, 1985]. Due to these experiences, error seeding is known to be very difficult to do.

Mills [1972] developed the technique of seeding errors to estimate the number of errors in a program. His method relied upon one member of a testing team seeding errors in a program and then having another member of the team attempt to find the seeded errors. Based upon the success of finding the seeded errors, it should be possible to estimate the total number of remaining unseeded errors in the program [Pfleeger, 2001].

Of course, if seeded errors are not reflective of real errors, then this method may give highly misleading results. Practice in software engineering has long shown that error seeding is very difficult. Human seeders may not consider certain kinds of errors that occur naturally [Meek & Siu, 1989]. Human seeders also may be subject to subconscious or conscious bias or preconceptions that could affect their error choices [Grigorjev, et. al, 2003; Meek & Siu, 1989]. Fundamentally, an attempt to produce errors by conscious effort that equate with errors produced unconsciously during code creation is very difficult [Knight and Ammann, 1985].

In general, it is very difficult to know a realistic artificial fault [Offutt & Hayes, 1996]. Humans tend to seed faults that are simpler to define and manage [Offutt & Hayes, 1996]. This may explain why error seeding methods, in practice, have successfully estimated total remaining errors for easier-to-find errors while being unsuccessful at estimating the residual number of difficult-to-find errors [Malaiya & Denton, 1998].

Grigorjev, *et al.* [2003] said that these difficulties require a useable taxonomy of error types as well as knowledge of their probabilities. Beyond this, there needs to be an understanding of how an error's position will influence its visibility and therefore its probability of detection.

Powell, *

## 2.3 The Warren Anderson Study

A second study to study SSAP detection rates was done by Warren Anderson [2004]. Anderson used a variation of the Galletta, *et al.* [1997] spreadsheet, adding several additional parts. Anderson gave the spreadsheet a total of 15 seeded errors.

This study had two parts. In the first, Anderson applied 11 static inspection tools to the seeded spreadsheet. Similar to what was done in the Nixon and O'Hara, [2001] study, a tool was given a one if it passed, a zero if it failed, and proportionate values if it almost passed or almost failed.

In the second part, Anderson recruited 65 subjects to manually inspect the spreadsheet. These subjects had a mean of 6.8 years of spreadsheet experience and a mean weekly spreadsheet use of 5.3 hours. Instead of using Excel, they used a simulation of Excel so that Anderson could collect data.

Figure 1 shows the results. On average, humans were 37% correct, while audit tools averaged 27%. So humans were somewhat better. However, as Anderson had hypothesized, there were strong differences between the errors that humans were better at finding and errors that spreadsheet static analysis tools were better at finding.

**Figure 1: Results from the Anderson Study**

| Error | Humans | Audit Tools | Significance |
|---|---|---|---|
| Value is stored as text | 15% | 82% | Highly significant |
| Y2K problem | 2% | 36% | Highly significant |
| Formula adds instead of multiplies | 38% | 55% | |
| Formula adds the wrong cell | 42% | 55% | |
| Vlookup | 6% | 9% | |
| Incomplete range used | 52% | 55% | |
| Countif | 8% | 9% | |
| Sums values when should not | 17% | 0% | |
| Typo: 25 instead of C25 | 55% | 36% | |
| Typographical error in data | 32% | 0% | |
| Countif | 51% | 18% | Significant |
| Omits a variable | 58% | 18% | Significant |
| Range is incorrect | 57% | 9% | Significant |
| Sum counts additional cells | 57% | 9% | Highly significant |
| Operator precedence error | 57% | 9% | Highly significant |
| Mean | 37% | 27% | |

## 2.4 Perspective

These two studies seem to indicate that spreadsheets inspection programs seem to be reasonably effective at tagging seeded errors. However, the Anderson [2004] study confirmed that the effectiveness of error flagging differed from 0% to 82% for different types of seeded errors. Yet neither study met good practices requirements for error seeding in software development—most importantly, the requirement to base seeded errors on a taxonomy of known errors and their relative frequencies and the requirement to seed errors on the basis of visibility. In particular, neither had many seeded domain knowledge errors or logic that was required but was omitted from the program.

## 3 THIS RESEARCH

The potential for spreadsheet analysis programs may or may not be very bright. The problem is that research has been done on seeded errors, and good error seeding is extremely difficult to do without having a major impact on results.

Consequently, we conducted a research project to examine human error detection versus SSAP error flagging, based on a corpus of spreadsheets developed by human subjects and therefore containing human-generated (natural) errors instead of seeded errors.

## 3.1 The Corpus

The corpus of 75 spreadsheets used in the study was created previously by undergraduate business students in an introductory junior-year management information systems course [Panko, 2000]. All had previously taken two semesters of accounting and an introductory computer course that taught Excel and other computer literacy tools. In their current course, they had undergone Excel refresher training.

The task was to create a two year pro-forma (projected) income statement from a word description. The task was called Kooker because the company being modeled made microwave slow cookers. This task levered the subjects' training in both accounting and Excel.

Each spreadsheet had a unique solution, and would have been almost impossible to get the bottom line figures if there were any errors in the spreadsheets. Error counting was done with the almost universally used original sin method. This method records errors only in cells where they occur. It does not count cells that were subsequently incorrect because of the error. Also, an error that was the same in both years of the analysis was treated as a single error.

## 3.2 The Taxonomy

The study used the Panko and Aurigemma [2010] error taxonomy that has been accepted for publication in *Decision Support Systems*. This taxonomy is based on human error research in a variety of fields. Figure 2 on the next page shows the full taxonomy. This study only considers quantitative errors, and it only goes down to the next level of planning versus execution errors. Although data were collected for lower-level errors, the number of these lower-level errors in most categories was too small to give meaningful results.

If an error is made before the developer begins to make an entry in a cell, this is a planning error. In contrast, if the developer makes an error when entering the planned formula or number in a cell, this is an execution error. The distinction is important in error detection because planning errors are less likely to leave an artifact on a spreadsheet than execution errors.

### 3.3 The SSAP-Aided Error-Finding Study

The first part of the current study was conducted as a final exercise in a class on spreadsheet development and management. Subjects were juniors and seniors in management information systems. By this time in the course, subjects had developed extensive Excel 2007 skills and spreadsheet development skills, including inspection skills.

The subjects were each required to apply Microsoft Error Check and Spreadsheet Professional (student version) to 50 spreadsheets in the corpus. Subjects in this experiment were told which cells contained errors in each spreadsheet. They were tasked to determine whether the Error Check and Spreadsheet Professional correctly flagged each error for further checking. (These programs do not find errors but rather flag a cell or region as meriting further attention.) If the subject determined that an error had been tagged, he or she decided whether the flagging would have been useful to an inspector. This required judgment. Subjects were told to give the software checker the benefit of the doubt—that is, to say that the flagging was useful if it was even slightly likely to be useful. This study, in other words, had a bias in favor of Error Check and Software Professional.

**Figure 2: Panko and Aurigemma Taxonomy**

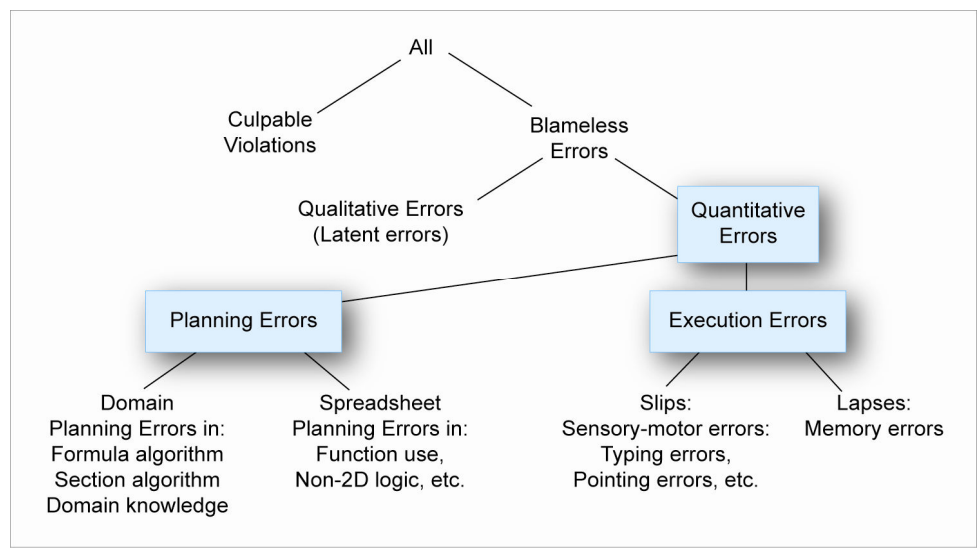

Source: Panko and Aurigemma [2010].

Although subjects attempted to run the two programs by themselves over the 50 spreadsheets, six spreadsheets had too many rows for the student version of Spreadsheet Professional because the developer had used extensive vertical white space.

Consequently, our analysis is limited to the 44 remaining spreadsheets. These spreadsheets contained 97 errors.

## 3.4 The Human Error Detection Study

In the second study, subjects were like those in the corpus development exercise, although they were taking the introductory MIS course in a later semester.

In the experiment, each subject was given *a single spreadsheet* from the corpus to inspect and look for errors. (They also did unrelated tasks in the experiment afterward.) Subjects were not given prior knowledge of whether their spreadsheet contained errors or if so, how many errors it contained. They were simply told that the spreadsheet might or might not be correct and that if it had errors, it might have one or more errors. Subjects recorded the cell or cells containing each error they discovered, gave a description of each error, and described how they would fix each error. They did not actually fix the errors.

## 4 RESULTS

Figure 3 gives the results of the two studies. The differences are stark. Unaided error inspection was able to detect 54 of the 97 errors. This 56% success rate is similar to detection rates in studies of inspection with seeded errors [Panko, 2010b].

**Figure 3: Results of the Two Studies**

| Error Category | Number | Detection for All Errors | Detection for Mistakes v Execution Errors |
|---|---|---|---|
| **Total Errors** | **97 (100%)** | | |
| Human Detections | 54 | 56% | |
| Software Detections | 5 | 0.52% | |
| **Mistakes** | **81 (84%)** | | |
| Human Detections | 39 | | 48% |
| Software Detections | 2 | | 0.25% |
| **Execution Errors** | **16 (16%)** | | |
| Human Detections | 11 | | 69% |
| Software Detections | 3 | | 1.88% |

Note: 4 errors could not be reliably classified as planning or execution errors.

In contrast, software-aided detection performed dismally. Only five errors were recorded as being usefully flagged—all by a single subject. None of these five detections was made by both tools. The number of opportunities for detection was 97 errors times five error seekers, times two programs per error seeker—a total of 970 opportunities to flag an error. So the 5 errors classified as being helpfully flagged were a mere 0.52% of all errors. For these natural human errors, then, software inspection programs were almost useless. They caught a mere one in 200 errors.

One reason for the poor showing of the SSAP tools may be that 84% of the errors were planning errors, while only 16% were execution errors. (Four errors could not be reliably classified as mistakes or execution errors and were not analyzed.)

In planning errors, which are also called mistakes, the person has the wrong plan before entering contents into a cell. Planning errors take place in the developer's head and may not leave an artifact on the spreadsheet for a spreadsheet static analysis tool to identify. Human beings also found mistakes difficult to identify. They found only 48% of these errors. In contrast, the SSATs did a correct detection in only 0.25% of all opportunities.

Execution errors, in which the person fails to carry out the planned cell entry accurately, are more likely to leave artifacts on the spreadsheet for the flagging functions to find. For execution errors, the success rate for SSATs did rise to 1.88%. However, this is still terrible. Human error detection also rose compared to error seeking with mistakes. Humans were able to find 69% of the errors.

Overall, the two spreadsheet static analysis tools performed terribly, and their detection strengths paralleled the detection strengths of people instead of compensating for weaknesses in human error detection.

## 5 DISCUSSION

In this study, attempts were made to find errors in a corpus of 44 spreadsheets previously developed by subjects from a word problem. These spreadsheets contained 97 errors. In the first part of the study, error-seeking was done by five subjects using the error-tagging functions of Excel Error Check and Spreadsheet Professional. These subjects knew the errors in the spreadsheet and merely had to determine if they were flagged by the software and how well the tagging indicated the error. In the second part of the study, 44 other subjects each examined a single spreadsheet from the corpus. The purpose of the study was to understand differences between the effectiveness of automated error flagging and human error inspection.

Previous studies of error detection have used spreadsheets seeded with errors by the experimenter. Error seeding has long been used in software development as a way to estimate the number of errors remaining in a program after testing. Software experience has shown that selecting seeded errors is extremely difficult to do and that errors must be selected on the basis of a good taxonomy of error types, known relative frequencies of occurrence, and knowledge of error visibility. Previous studies that seeded spreadsheets with errors did not meet these criteria. Consequently, this study used a corpus of humanly developed spreadsheets in order to understand error-seeking behavior for natural errors.

### 5.1 Limitations

The strength of this research is that it did not resort to error seeding. However, it did use undergraduate students rather than experts. In addition, the two studies do not constitute an experiment because subjects were not drawn randomly from the same population and assigned randomly to the two conditions.

To reduce the problems that this non-randomization may have caused, the effort was made to have possible biases skew the results in favor of the Excel static assessment tools. First, the students involved in the human search from errors were taken from an introductory third-year computer course on information systems and came from many majors. In contrast, the use of static analysis tools was conducted by seniors in information systems who were completing a full-semester course in spreadsheet development and testing. Second, while the human inspection subjects did not know where the errors were in the spreadsheets they checked, the static analysis subjects knew exactly where the errors were.

## 5.2 Findings

Overall, the 33 human inspectors found 56% of the 97 errors, while only 0.52% errors were tagged by Error Check and Spreadsheet Professionals only 0.52% of the time. Human subjects detected only 48% of the errors, while the software programs found only 0.25%.

If these percentages occur in other contexts, error tagging might not even be a good way of doing prescreening spreadsheets, much less replacing human error inspection.

## 5.3 Why So Bad Compared to Prior Studies of Seeded Errors?

In previous studies using seeded errors, error inspection software did far better, finding 50% to 80% of all errors. Why did they perform so poorly in our study? One plausible reason is that we only looked at error flagging. Programs such as Spreadsheet Professionals have other facilities for finding errors, including map functions that visually portray patterns of similarity and dependency in the spreadsheet. We did not test these capabilities formally on our corpus of errors.

The other possibility is that natural errors do not look like the seeded errors used in previous studies. In previous studies, most of the errors were things that spreadsheet static analysis tools were designed to find. In contrast, in our study, 84% of the errors were planning errors, which occurred within the developer's heads. Even our execution errors did not look much like the errors that Error Check and Spreadsheet Professional error tagging were created to find.

## 5.4 The Powell, Baker, and Lawson, (2008) Study

Powell, Baker, and Lawson [2008] developed a methodology for the inspection of spreadsheet models. In their approach, the inspector first ran two error inspection programs over the spreadsheet and then did human inspection of all remaining formulas.

For their first 50 spreadsheets, they reported what percentage of errors were detected by each tool. Figure 4 shows that the automated tools caught all but 17.8% of the errors found in the study. So in this case, automated tools worked far better than human inspection.

**Figure 4: Stages in Which Errors are Found**

| Tool | Percent |
|---|---|
| Map Analysis | 43.0% |
| Spreadsheet Professional Tests | 35.0% |
| Human Code Inspection | 17.8% |
| XL Analyst | 2.0% |
| **Total** | **100.00%** |

Source: Powell, Baker, and Lawson (2008)

There are several plausible reasons for the differences between our study and theirs. An obvious explanation is that the human errors in our corpus were unrepresentative of human errors in general. After all, the 50 spreadsheets inspected by the Powell, Baker, and Lawson (2008) study were not just developed by humans. They were operational spreadsheets developed by real business developers. Our spreadsheets were created by undergraduate business students. In addition, by not using the mapping function, we hobbled the software inspection programs.

On the other hand, it may be that the Powell, Baker, and Lawson [2008] methodology under-detected errors, especially the type of errors that humans make and which spreadsheet inspection programs cannot find. There are two reasons to consider under detection is plausible.

In software code inspection, it is only possible to inspect 100 to 200 lines of code per hour [Fagan, 1972]. When this rate is succeeded, the detection rate falls rapidly [Panko, 2010a]. For the 50 spreadsheets in the Powell, Baker, and Lawson [2008] corpus, the inspectors only spent a mean of only 3.25 hours per spreadsheet, including the running of Spreadsheet Professional Tests and XL Analyst and the recording and classification of errors. Their methods require the inspector to look at all remaining formulas, at least briefly. However, given human error research on error detection, the inspectors may not have had enough time to do a good job of this. Powell, Baker, and Lawson [2008] defending the relatively brief period of inspection by saying that, "This seemed to be a reasonable amount of time to devote to auditing a spreadsheet of importance to an organization." They did not elaborate.

Also, for the 50 spreadsheets used in this study, the inspectors were simply given the spreadsheets. There was no discussion with the spreadsheet authors about the objectives of the spreadsheets or of the domain knowledge concepts required to build the spreadsheet. This plausibly reduce the inspector's ability to find planning errors or to understand execution errors. In later studies, the methodology was expanded to include initial preparation, but not having a domain understanding of each spreadsheet may plausibly have reduced the inspectors' ability to find planning errors.

Powell, Baker, and Lawson [2008] never suggested that their method could find all errors. However, if it seriously under-counted planning errors and many types of execution errors, the low percentage of detections made by human inspection is seriously misleading.

## 5.5 Next Steps

Given the expense of comprehensive inspection and the performance of spreadsheet inspection programs in the Powell, Baker, and Lawson [2008] study and in studies using seeded errors, it would be good if spreadsheet inspection programs could greatly reduce the cost of doing inspections. However, studies that used seeded errors did not do careful error seeding, and the Powell, Baker, and Lawson study was never verified or validated to estimate what percentage of errors it found. Our study found that error-tagging features in two spreadsheet inspection programs did very poorly compared to human inspectors in a sample of 44 spreadsheets created by previous subjects.

Given the importance of the issue and the disparity of results so far, we need to do considerably more research in this area. We need to have a corpus of operational spreadsheets or unit spreadsheets during development. We then need to run software inspection programs against these spreadsheets and also do full Fagan [1976] code inspection against the same spreadsheets. For the code inspection part of the study, the fact that multiple inspectors will independently review the spreadsheets allows the estimation of what fraction of all errors of various types remained undetected.